\documentclass[twocolumn]{aastex631}
\newcommand{\rpm}{\sbox0{$1$}\sbox2{$\scriptstyle\pm$}
  \raise\dimexpr(\ht0-\ht2)/2\relax\box2 }
\usepackage{hyperref}
\usepackage{float}
\usepackage{textgreek}

\begin{document}

\title{Inferring Stellar Parameters from Iodine-Imprinted Keck/HIRES Spectra with Machine Learning}

\author[0000-0003-4705-3006]{Jude Gussman}
\affiliation{Department of Astronomy, Indiana University, Bloomington, IN 47405, USA}

\author[0000-0002-7670-670X]{Malena Rice}
\affiliation{Department of Astronomy, Yale University, New Haven, CT 06511, USA}
\email{malena.rice@yale.edu}


\begin{abstract}
    The properties of exoplanet host stars are traditionally characterized through a detailed forward-modeling analysis of high-resolution spectra. However, many exoplanet radial velocity surveys employ iodine-cell-calibrated spectrographs, such that the vast majority of spectra obtained include an imprinted forest of iodine absorption lines. For surveys that use iodine cells, iodine-free ``template'' spectra must be separately obtained for precise stellar characterization. These template spectra often require extensive additional observing time to obtain, and they are not always feasible to obtain for faint stars. In this paper, we demonstrate that machine learning methods can be applied to infer stellar parameters and chemical abundances from iodine-imprinted spectra with high accuracy and precision. The methods presented in this work are broadly applicable to any iodine-cell-calibrated spectrograph. We make publicly available our spectroscopic pipeline, the Cannon HIRES Iodine Pipeline (CHIP), which derives stellar parameters and 15 chemical abundances from iodine-imprinted spectra of FGK stars and which has been set up for ease of use with Keck/HIRES spectra. Our proof-of-concept offers an efficient new avenue to rapidly estimate a large number of stellar parameters even in the absence of an iodine-free template spectrum.
\end{abstract}

\section{Introduction} 

One of the most common techniques used to search for and characterize exoplanet systems involves measuring the Doppler shift of a stellar spectrum over time. This time-varying Doppler shift encodes the line-of-sight reflex motion of a star  as it orbits around the center of mass of a planetary system. This technique, known as the ``radial velocity'' (RV) method, enabled the first discovery of an exoplanet orbiting a Sun-like star \citep{mayor1995jupiter}. Since then, precision RV measurements have been applied to study thousands of planetary systems.

To obtain sufficiently precise Doppler shift measurements for exoplanet detection, spectrographs require an extremely well-calibrated wavelength solution that can be used as a reference to calibrate the observed shifts. Some spectrographs (e.g. Keck/HIRES \citep{vogt1994hires, valenti1995determining, butler1996attaining}, APF \citep{vogt2014apf}, HET/HRS \citep{cochran2004first}, and PFS/Magellan \citep{crane2006carnegie, crane2008carnegie, crane2010carnegie}) -- referred to hereafter as ``iodine-cell spectrographs'' -- employ an iodine cell that imprints molecular iodine absorption lines onto each stellar spectrum to provide a precise wavelength reference for the Doppler shift. These surveys determine the RV profile of a star by obtaining many iodine-imprinted spectra, with Doppler shifts that encode information about the mass and orbital properties of neighboring companions.

Radial velocity surveys with iodine-cell spectrographs typically obtain one or more high-resolution iodine-free spectra, known as ``template" spectra, that are analyzed to determine important stellar parameters such as effective temperature ($T_{\rm eff}$), the projected stellar rotation speed ($v\sin i$), surface gravity ($\log g$), and metallicity ([Fe/H]) \citep[e.g.][]{valenti2005spectroscopic}, as well as elemental abundances within the stellar atmosphere \citep{brewer2016spectral, brewer2018spectral}. However, observations of these template spectra are relatively time-intensive, and template spectra cannot always be obtained for faint stars \citep{dalba2020multiple}.

Machine learning offers a compelling new avenue to derive extensive spectral information even from relatively noisy input spectra. \textit{The Cannon} \citep{ness2015cannon}, a widely used data-driven machine-learning spectroscopy program, trains a flexible model at each wavelength bin from a set of continuum-normalized spectra with known stellar parameters -- hereafter used interchangeably with the terms ``labels'' or ``stellar labels'' in the context of \textit{The Cannon} -- then applies this model to extrapolate the stellar labels associated with new input spectra. Furthermore, \textit{The Cannon} reliably transfers labels across complex models trained with 17-18 stellar labels, enabling the inference of multiple abundances in addition to key global stellar parameters \citep{casey2016cannon, rice2020stellar}.

In this paper, we demonstrate that stellar parameters can be accurately derived from iodine-imprinted radial velocity spectra, in spite of the elevated noise added by the superimposed iodine lines. Our proof-of-concept shows that 18 stellar parameters and abundances (T$_{\rm eff}$, $\log g$, $v\sin i$, [Fe/H], [C/H], [N/H], [O/H], [Na/H], [Mg/H] ,[Al/H], [Si/H], [Ca/H], [Ti/H], [V/H], [Cr/H], [Mn/H], [Ni/H], and [Y/H]) can be inferred from iodine-imprinted Keck/HIRES spectra using \textit{The Cannon}. 

We also describe our code, the Cannon HIRES Iodine Pipeline (CHIP), and make it available for public use. CHIP accepts iodine-imprinted stellar spectra, and it streamlines the stellar parameter inference process for Keck/HIRES spectra by retrieving and fully reducing Keck/HIRES iodine-imprinted RV observations from the Keck Observatory Archive (KOA). The automated spectral reduction process involves continuum normalization, cross-correlation, and interpolation. After the reduction process is complete, CHIP inserts the reduced spectra into \textit{The Cannon} for parameter inference.

We first describe our dataset in Section \ref{section:dataset}. Then, we discuss the CHIP pipeline and the associated spectral reduction process in Section \ref{section:CHIP}. Our model performance, demonstrated with a sample test set of Keck/HIRES stars, is summarized in Section \ref{section:results}. Finally, we conclude in Section \ref{section:conclusions}.

\section{Dataset}
\label{section:dataset}

To train and test our model, we leveraged a dataset consisting of (1) iodine-imprinted radial velocity spectra and (2) the associated pre-determined stellar labels used to verify the performance of our code. Our stellar sample consists of 1201 F, G, and K dwarf stars drawn from the \citet{brewer2016spectral} Spectral Properties of Cool Stars (SPOCS) catalog, with the same sample cuts applied as in \citet{rice2020stellar}. These stars were originally chosen for their suitability for high-resolution spectroscopy with the Keck/HIRES instrument. The labels for all stars within this sample were determined in \citet{brewer2016spectral} through a spectral synthesis modeling procedure that leveraged the Spectroscopy Made Easy (SME) software \citep{valenti1996spectroscopy} to analyze high-resolution, iodine-free Keck/HIRES spectra. 

After defining our initial sample of F, G, and K dwarf stars, we queried the KOA to obtain iodine-imprinted spectra for each star using the publicly available NASA Exoplanet Science Institute HIRES Precision Radial Velocity (NExScI HIRES PRV) pipeline \citep{butler1996attaining, howard2010planetsurvey}.\footnote{\url{https://caltech-ipac.github.io/hiresprv/index.html}}. To maximize the quality of our derived stellar parameters, our sample included only the highest SNR spectrum for each star.

\begin{figure*}
    \centering
    \includegraphics[width=0.85\textwidth]{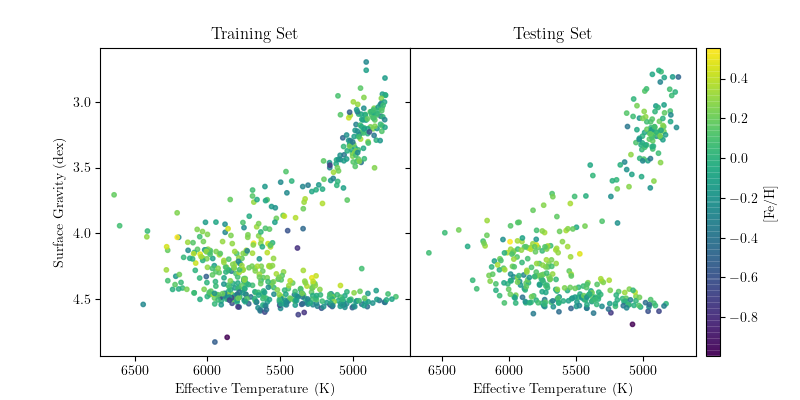}
    \caption{A comparison of the stellar properties spanned by the training and test sets. The left diagram depicts the 60\% of stars in the training set, while the right diagram showcases the remaining 40\% in the test set.}
    \label{fig: HRtraintest}
\end{figure*}

To verify the performance of our methods, we divided our dataset into two parts: a training set and a test set. The training set included 60\% of the full dataset, while the remaining 40\% was reserved for the test set. Figure \ref{fig: HRtraintest} compares the properties of stars in the training and test sets, showing that they span a similar parameter space. The foundational dataset for our analysis, which draws from Tables 8 and 9 in \citet{brewer2016spectral} and which includes columns labeled `HIRESID' and the stellar parameters `TEFF', `LOGG', `VSINI', `CH', `NH', `OH', `NaH', `MgH', `AlH', `SiH', `CaH', `TiH', `VH', `CrH', `MnH', `FeH', `NiH', and `YH', is accessible on our GitHub repository as \href{https://github.com/jgussman/CHIP/blob/main/data/spocs/stellar_parameters.csv}{stellar\_parameters.csv}.

\section{The Cannon HIRES Iodine Pipeline (CHIP)}
\label{section:CHIP}

\subsection{Overview of the pipeline}

The Cannon HIRES Iodine Pipeline (CHIP) is a comprehensive, automated data processing and analysis pipeline presented in this work. CHIP derives accurate stellar parameters from iodine-imprinted Keck/HIRES spectra, utilizing the machine learning algorithm \textit{The Cannon}. The pipeline takes as inputs a set of reduced, iodine-imprinted spectra; the associated pixel-by-pixel inverse variance $1/\sigma^2$; and the known stellar parameter values. 

CHIP streamlines the spectral reduction process by retrieving and fully reducing iodine-imprinted RV observations from the KOA. The reduction process implemented by CHIP includes essential steps such as continuum normalization, cross-correlation, and interpolation. Upon completion of the reduction process, CHIP saves the reduced fluxes, inverse variances, and wavelengths for each spectrum. 

Ideally, the user should visually inspect at least some subset of the reduced data at this stage to verify that the cross-correlation process has consistently aligned all spectra -- particularly when examining low-SNR datasets for which the cross-correlation process may be less accurate. This alignment is crucial: \textit{The Cannon} processes spectra on a pixel-by-pixel basis and does not, therefore, consider full spectral features. As a result, small inconsistencies in the wavelength solution may result in significantly impaired performance. 

After the user verifies that the reductions were performed correctly, the data is fed into \textit{The Cannon} for training and parameter inference. The output is a set of extracted parameters for each star, along with their associated uncertainties. The automated nature of the pipeline enables uniform and rapid parameter extraction while maintaining a high level of accuracy.

CHIP uses the K-fold cross-validation method \citep[e.g.][]{rodriguez2009sensitivity, wong2019reliable} from the \texttt{scikit-learn.model\_selection} Python library\footnote{See \url{https://scikit-learn.org/stable/modules/cross_validation.html} for more details on the \texttt{scikit-learn} implementation of K-fold and other cross-validation methods.} to monitor model performance during training and to prevent over-fitting. The technique involves splitting the training set into \( K \) subsets or `folds'. The model is then trained on \( K-1 \) of these folds and validated on the remaining fold. This process is repeated \( K \) times, with each fold serving as the validation set exactly once. The test set is kept separate from the training and validation sets and is used to evaluate how well the best model performs on data that it did not train on. By using independent data for testing, it is possible to detect and address any issues that arose during training, such as over-fitting. Overall, this approach helps to ensure that trained models are accurate, reliable, and able to generalize well to new data.

CHIP includes the option to perform mini-batch training, which can improve the efficiency of the training process by breaking the training dataset into smaller subsets or batches. This can reduce the memory requirements for training and allow for faster convergence of the model. In addition, mini-batch training can be used to regularize the optimization process and reduce the effects of noisy data. While the results presented in this work do not leverage mini-batch training due to the size of our dataset, which has a relatively small memory footprint, this option allows for flexibility when considering very large-scale spectroscopic datasets. We note that it is important to choose an appropriate batch size that balances these benefits with the risk of overfitting or underfitting the data.

To assess the effectiveness of the trained models, CHIP computes a normalized mean difference for each of the 18 stellar parameters provided in the SPOCS catalog (including T$_{\rm eff}$, $\log g$, $v\sin i$, and 15 abundances). These parameters are inferred from the iodine-imprinted spectra, and CHIP selects the best-performing model for all the parameters based on the average evaluation score of the model during training. Notably, the chosen best model can reliably derive all 18 parameters from the iodine-imprinted spectral dataset, as we demonstrate in subsequent sections of this paper.

\begin{figure*}
  \centering
  \includegraphics[width=0.98\textwidth]{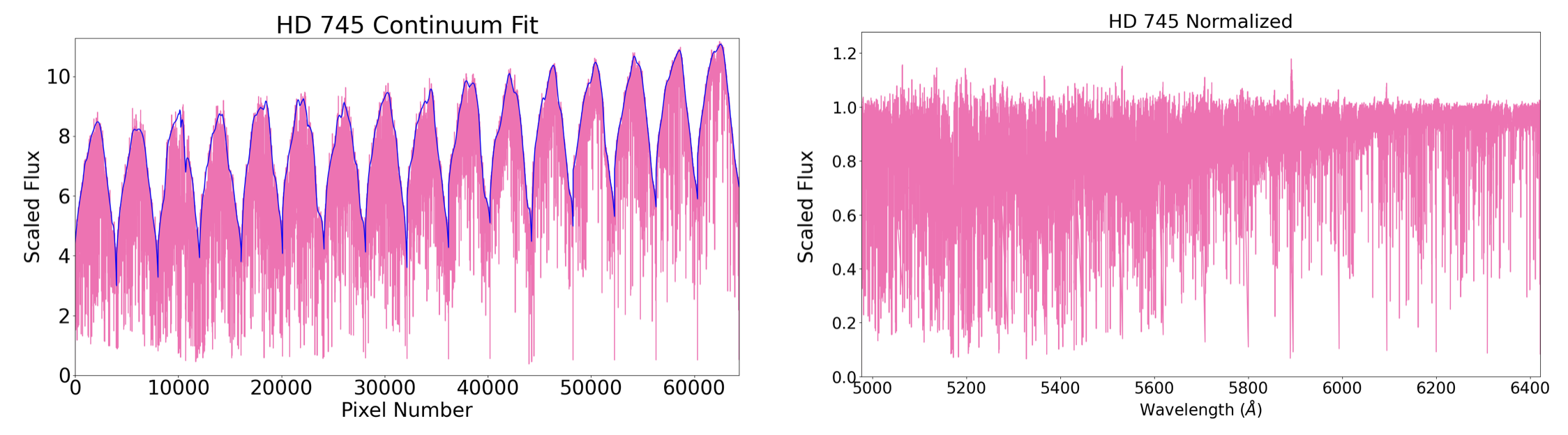}
\caption{\textit{Left:} HD 745's AFS continuum normalization fit (blue) over-plotted on top of the deblazed spectrum (pink). \textit{Right:} HD 745 spectrum after AFS continuum normalization.}
\label{fig:continuumfit}
\end{figure*}

\subsection{Data retrieval from the NExScI PRV pipeline}
While CHIP is also able to accept custom datasets provided by the user, for ease of use the code includes the functionality to directly extract deblazed RV observations from the NExScI HIRES PRV pipeline. For our purposes, we required only a single iodine-imprinted spectrum for each star of interest. 


In the data acquisition stage, CHIP downloads all available iodine-imprinted spectra in the KOA corresponding to each targeted star. Subsequently, CHIP calculates the signal-to-noise ratio (SNR) using the median photon counts from the fifth echelle order of each spectrum ($\sim$5273-5367 \,\AA; the middle of the iodine-imprinted region) to identify the spectrum that offers the highest SNR, adopting a gain value \(G=2.09\) for HIRES. In cases where none of the spectra satisfy the requirement SNR $>100$, the star is deemed unsuitable for inclusion in the dataset, ensuring a robust and high-quality data sample. We emphasize that this SNR requirement is specifically necessary for the \textit{training set}, but not for a general test set -- that is, \textit{The Cannon} models trained with high-SNR data have been shown to accurately infer high-precision labels from significantly lower-SNR spectra than those used in the training set \citep{ness2015cannon, casey2016cannon}.



\subsection{Continuum normalization}
A spectrum must be continuum-normalized before being fed into \textit{The Cannon}, such that the baseline flux of the spectrum is unity. By default, CHIP uses the Alpha-shape Fitting to Spectrum (AFS) continuum normalization method \citep{xu2019modeling}. Figure \ref{fig:continuumfit} demonstrates the continuum fit to the HD 745 system using the AFS method. This continuum normalization method reliably produces a flat continuum baseline, improving the accuracy of the subsequent cross-correlation step. 






\subsection{Rest frame wavelength alignment}

In CHIP, we leverage a specialized version of the Empirical SpecMatch (\texttt{SpecMatch-Emp}) algorithm exclusively for the purpose of wavelength shifting, aligning each spectrum to its rest wavelength. Originally presented in \citet{yee2017precision}, \texttt{SpecMatch-Emp} is designed for a broader range of functionalities, including spectral matching and stellar parameter estimation. Our adaptation of the code focuses solely on its capability for precise wavelength alignment. 

Our modified version retains \texttt{SpecMatch-Emp}'s bootstrapping approach for cross-correlation, a method proven effective for aligning spectra across diverse stellar types. In the original framework, a ladder of template spectra is employed to identify the best reference spectrum for each target, based on the largest median cross-correlation peak. This approach is applied through CHIP to shift each target spectrum to its rest wavelength.

\subsection{Interpolation onto a common wavelength grid}

After shifting all spectra to their rest wavelength frame, CHIP performs linear interpolation to resample all spectra onto a final, common wavelength grid. This uniformity in wavelength is crucial for subsequent machine learning analyses using \textit{The Cannon}, which compares spectra on a pixel-by-pixel basis. For spectral regions that overlap across echelle orders, only the first occurrence of each overlap region is retained, while the occurrences in higher echelle orders are removed. The final, fully reduced spectra and associated inverse variances for each spectrum are saved for input into \textit{The Cannon}.

\subsection{Evaluation}

CHIP applies an evaluation function that assesses the model's accuracy in recovering known parameters during the validation stage of model selection. The evaluation function is defined as the mean difference between the true and predicted values of the parameters for each star after standardizing all parameters with the \texttt{sklearn.preprocessing.StandardScaler} Python method. CHIP uses this evaluation function for each fold of the training and validation data splits, accruing a list of evaluation scores. The mean of these evaluation scores is computed to yield a single score that quantifies the model's accuracy across all folds.

When examining model performance across hyperparameters, CHIP's evaluation function may also be used to compare the performance of models. Hyperparameters that may be varied include, for example, the polynomial order of the model used by \textit{The Cannon}; label censoring; regularization; and the application of masks over certain spectral regions. We note that the results presented in this study were achieved using our default settings: a polynomial order of 2 and no censoring, regularization, or mask application.

After selecting a set of hyperparameters, a final model retaining these hyperparameters is trained on the combined training and validation datasets. This model is then applied to the test set to evaluate its generalizability.



\section{Results} 
\label{section:results} 

\begin{figure*}
  \centering
   \includegraphics[width=0.98\textwidth]{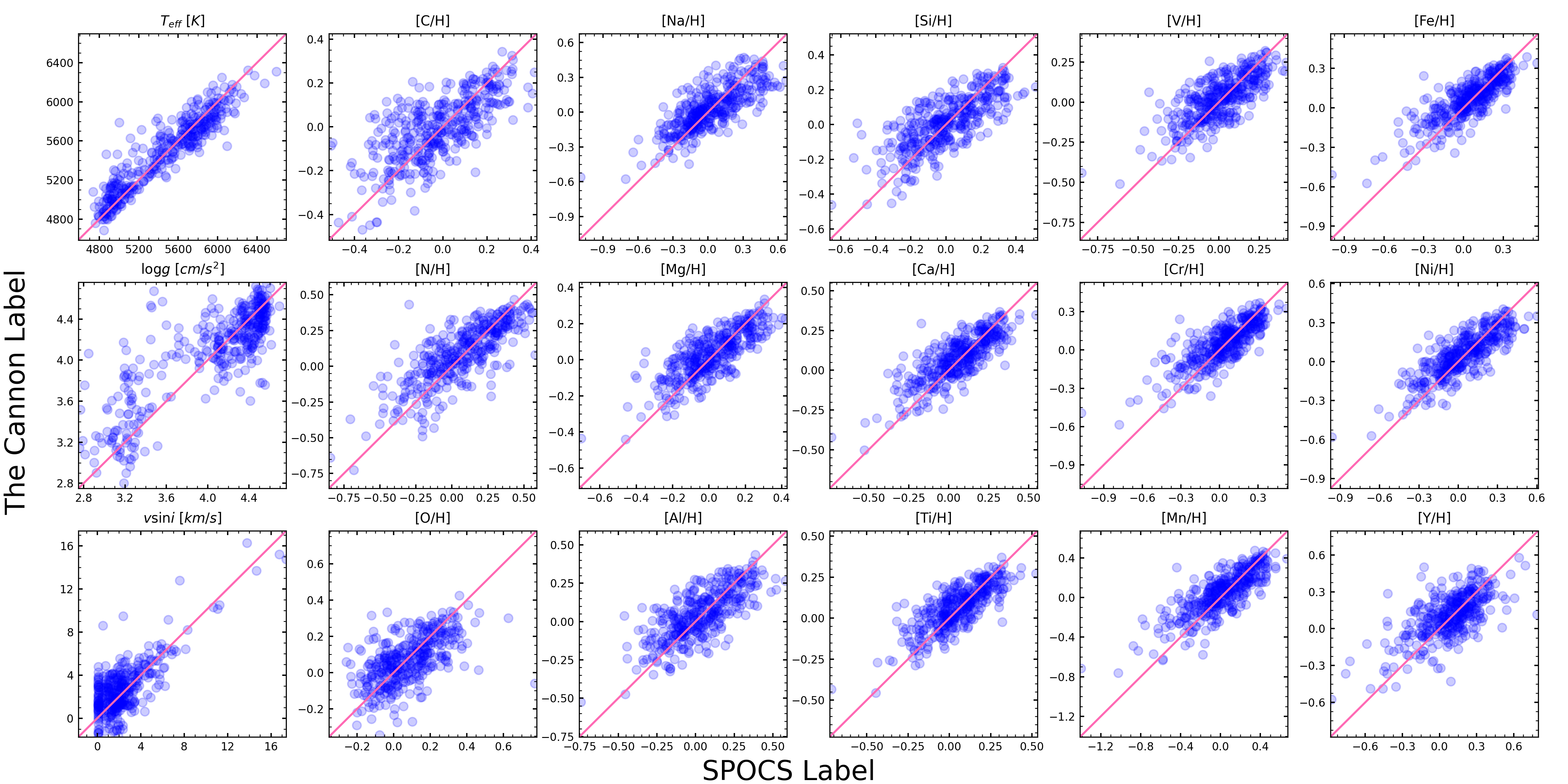}
\caption{Each blue point represents one of 372 stars in our test set. All 18 label values from the SPOCS catalogue are plotted against our trained model's predicted values. The pink line through each panel displays the line of equality. Three extreme outlier stars, which performed poorly across all labels, were removed from the sample prior to creating this figure; such outliers would be straightforward to identify in practice, since they correspond to nonphysical stellar properties.}
\label{fig:CannonvsSPOCS}
\end{figure*}


Examining our dataset with CHIP, we obtained the test set results shown in Figure \ref{fig:CannonvsSPOCS}. Our final model employs 2-fold cross-validation, and a 2nd-degree polynomial function is used across labels for the generative model within \textit{The Cannon}.



Figure \ref{fig:CannonvsSPOCS} compares our top-performing model against the ``true'' SPOCS catalogue labels using the test set. For models that generalize well, the results should be most closely aligned with the one-to-one line that corresponds to exact label recovery. The fidelity of our results can also be seen in Table \ref{tab:table_results}, which contains the uncertainties and mean difference between the SPOCS catalogue references. 

As expected, CHIP's uncertainties are larger than those of previous works that leveraged iodine-free spectra \citep[e.g.][]{brewer2016spectral, brewer2018spectral, teske2019do, rice2020stellar}. The derived CHIP uncertainties are typically about a factor of 10 larger than the uncertainties from direct spectral modeling in \citet{brewer2016spectral}, and a factor of 2-3 larger than uncertainties from \textit{The Cannon} applied to iodine-free spectra in the same systems. However, we still find a clear correlation between the SPOCS labels and our derived parameters. We note that our final model has a tendency to slightly overpredict the values of most of our labels, as shown by the mean differences in Table \ref{tab:table_results}.

Our derived uncertainties are generally comparable to systematic differences observed across stellar parameter inference techniques \citep{hinkel2016comparison}. We note that some parameters -- for example $v\sin i_*$ and $\log g$ -- are poorly behaved in our recovery. This suggests that the imprinted iodine impacts the apparent line widths derived from \textit{The Cannon}, and signal pollution in the wings of lines adds scatter to the final solution. Overall, the performance of the model on the test set (see Figure \ref{fig:CannonvsSPOCS} and Table \ref{tab:table_results}) demonstrates a clear ability to accurately recover most stellar parameters of interest -- albeit with larger uncertainties than obtained from physical models -- even in the presence of imprinted iodine lines.

\begin{deluxetable}{lcccc}
\tablecaption{Test results from our final ensemble models, which employed 2-fold cross-validation with a 2nd-degree polynomial function used to model labels within \textit{The Cannon}. Label uncertainties from CHIP are denoted as $\sigma_{\rm CHIP}$, while $\sigma_{\rm Rice}$ and $\sigma_{\rm Brewer}$ are the parameter uncertainties reported in \citet{rice2020stellar} and \citet{brewer2016spectral}, respectively. The mean difference is given as the mean of the (CHIP - SPOCS) values for each parameter. The three extreme outlier stars that were removed from Figure \ref{fig:CannonvsSPOCS} were also excluded before deriving these metrics.}
\label{tab:table_results}
\tabletypesize{\scriptsize}
\tablehead{
\colhead{Label} & \colhead{Mean difference} & \colhead{$\sigma_{\rm CHIP}$} & \colhead{$\sigma_{\rm Rice}$} & \colhead{$\sigma_{\rm Brewer}$} }
\tablewidth{\linewidth}
\startdata
     $T_{\rm eff}$  &                            16 &                          156 &     56 &      12.5  \\ \hline
  $\log g$ &                             0.048 &                            0.296 &      0.09 &       0.014 \\ \hline
  $v \sin i$  &                             0.374 &                            1.650 &      0.87 &       0.350 \\ \hline
      C/H &                            0.015 &                            0.131 &      0.05 &       0.013 \\ \hline
      N/H &                             0.023 &                            0.158 &      0.08 &       0.021\\ \hline
      O/H &                            -0.008 &                            0.135 &      0.07 &       0.018 \\ \hline
     Na/H &                             0.010 &                            0.162 &      0.05 &       0.007 \\ \hline
     Mg/H &                             0.015 &                            0.103 &      0.04 &       0.006 \\ \hline
     Al/H &                             0.015 &                            0.136 &      0.04 &       0.014 \\ \hline
     Si/H &                             0.009 &                            0.131 &      0.03 &       0.004\\ \hline
     Ca/H &                             0.020 &                            0.109 &      0.03 &       0.007 \\ \hline
     Ti/H &                             0.016 &                            0.101 &      0.04 &       0.006 \\ \hline
      V/H &                             0.016 &                            0.119 &      0.06 &       0.017 \\ \hline
     Cr/H &                             0.025 &                            0.128 &      0.04 &       0.007 \\ \hline
     Mn/H &                             0.033 &                            0.170 &      0.05 &       0.010 \\ \hline
     Fe/H &                             0.025 &                            0.119 &      0.03 &       0.005 \\ \hline
     Ni/H &                             0.019 &                            0.126 &      0.04 &       0.006 \\ \hline
      Y/H &                             0.022 &                            0.174 &      0.08 &       0.015 \\ \hline
\enddata
\end{deluxetable}

\section{Conclusions}
\label{section:conclusions} 

In this work, we demonstrate that stellar parameters can be derived from iodine-imprinted spectra through machine learning methods. We present an algorithm, CHIP, that can estimate a total of 18 stellar parameters and chemical abundances from iodine-imprinted stellar spectra using the SPOCS catalogue as a training set. While the associated uncertainties in derived parameters using CHIP are substantially larger than those from spectral fitting algorithms leveraging iodine-free spectra (see Table \ref{tab:table_results} for detailed comparisons), they are comparable to systematic differences across different parameter inference methods \citep{hinkel2016comparison}. Our work offers a proof-of-concept for parameter derivation even from systematically contaminated spectra.

For ease of usability, CHIP has been designed with in-built functionality to derive and reduce Keck/HIRES spectra drawn from the KOA. However, the same algorithm can, in principle, be applied to any iodine-imprinted spectral dataset. Further details about the code and its implementation can be found on the \href{https://github.com/jgussman/CHIP}{CHIP} GitHub repository.

\section{Acknowledgements} 
We thank John Brewer for helpful discussions, and Songhu Wang for his generous support in providing access to Indiana University's high-performance computing resources during this project. We thank the anonymous reviewer for comments that have improved this manuscript. This research was supported in part by Lilly Endowment, Inc., through its support for the Indiana University Pervasive Technology Institute. All models described in this work were trained and tested on Indiana University's large memory computer cluster, Carbonate. 

M.R. acknowledges support from Heising-Simons Foundation Grant \#2023-4478. The data leveraged in this work were obtained at the W. M. Keck Observatory, which is operated as a scientific partnership among the California Institute of Technology, the University of California and the National Aeronautics and Space Administration. The Observatory was made possible by the generous financial support of the W. M. Keck Foundation. This research has made use of the Keck Observatory Archive (KOA), which is operated by the W. M. Keck Observatory and the NASA Exoplanet Science Institute (NExScI), under contract with the National Aeronautics and Space Administration.

\software{\texttt{numpy} \citep{oliphant2006guide, walt2011numpy, harris2020array}, \texttt{matplotlib} \citep{hunter2007matplotlib}, \texttt{pandas} \citep{mckinney2010data}, \texttt{scikit-learn} \citep{scikit-learn}, \texttt{scipy} \citep{virtanen2020scipy}, \texttt{SpecMatch-Emp} \citep{yee2017precision}, \textit{The Cannon} \citep{ness2015cannon},}

\bibliography{bibliography}
\bibliographystyle{aasjournal}

\end{document}